# Effects of Wax Impregnation on Contact Resistivity between REBCO Tapes

Jeremy W. Levitan, Jun Lu, Brent Jarvis, and Hongyu Bai

*Abstract*— Advances on no-insulation REBCO coil technology has made understanding and controlling contact resistivity increasingly important. Praffin (wax) impregnation is a process that has been used for improving mechanical stability of insulated and no-insulation REBCO coils. Wax impregnation is beneficial in both no-insulation coils and insulated coils with additional copper stabilizer or multiple conductors. In the latter scenario, contact resistance between conductor and additional stabilizer is also important. It is crucial to understand the effects of wax impregnation on contact resistivity ($R_{ct}$). We designed and built an apparatus to use short REBCO samples which simulates the behavior of $R_{ct}$ in a pancake coil during the wax impregnation process. $R_{ct}$ was measured at 77 K before and after the wax impregnation. In addition, a single pancake coil was wound to test the effect of wax impregnation. This coil simulates the NHMFL 32 T magnet Coil A in winding stresses. $R_{ct}$ was measured at 77 K and 4.2 K before and after wax impregnation. We found that wax impregnation does not significantly change contact resistivity. This means that wax impregnation can be used in coils without compromising the current sharing ability between turns. The experimental process and results are discussed.

## I. Introduction

Advances on no-insulation (NI) REBCO coil technology [1],[2] have made understanding and controlling contact resistivity ($R_{ct}$) increasingly important. $R_{ct}$ is the critical parameter that controls a coil's charging and quench behaviors, and is unique to a NI coil. The control of $R_{ct}$ has been studied [3]-[7].

Paraffin (wax) impregnation is a process that can improve a REBCO coil's mechanical stability and has been used for insulated [8] and the layer-no-insulation [9] REBCO coils. It is conceivable that the impregnation with wax, an insulator, may have impact on the contact resistivity of a NI coil. The effect of wax impregnation on $R_{ct}$ of the layer-no-insulation coil is not discussed in ref [9]. If wax impregnation were to be used in an NI coil, it is necessary to study its impact on $R_{ct}$. In addition, it is important to understand the effects of wax impregnation on $R_{ct}$ in a REBCO coil wound with two-in-hand conductors or conductor with additional copper stabilizer, which relies on transverse contact for current sharing. In this work, short sample $R_{ct}$ tests using REBCO/copper/REBCO stacks were performed before and after wax impregnation to determine the effects of the impregnation process on $R_{ct}$. In addition, a no-insulation single pancake coil was wound with its turn-to-turn contact pressure comparable to that of the NHMFL 32 T Coil A [8]. The decay time constant of this coil was measured at 77 K and 4.2 K before and after wax impregnation. The $R_{ct}$ values were analyzed and presented.

## II. Experimental Methods

### A. Short Sample Test

An apparatus was designed and fabricated to hold samples at a constant contact pressure during the wax impregnation as well as during $R_{ct}$ measurement at 77 K, as shown in Figure 1. The contact pressure was applied to the REBCO/Cu/REBCO stack by two stainless-steel screws. The torque on the screw versus contact pressure was calibrated using pressure sensitive film (Fuji film) whose color indicates pressure and pressure uniformity. This was done by replacing the Cu tape with a strip of Fuji film. Belleville washers were used on the screws which ensured that pressure on the REBCO was maintained during the impregnation at elevated temperature and during $R_{ct}$ testing at 77 K. This was to account for differences in thermal expansion

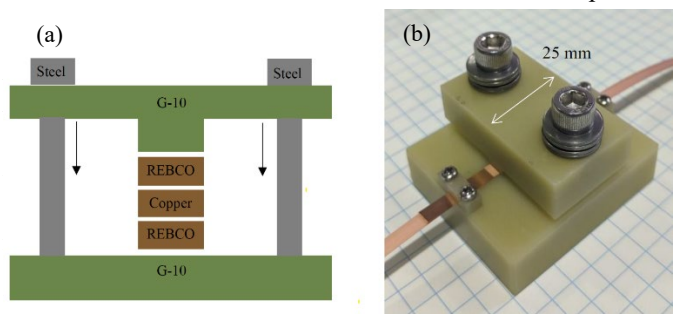

Fig. 1: (a) Schematic of the apparatus. REBCO tape and a 125 μm thick, pure Cu tape were stacked. The contact pressure is applied by two stainless steel screws (b) a photo of the apparatus.

between the stainless-steel screws, the G-10 parts, and the samples. The REBCO conductor is SuperPower SCS4050-AP with

This work is supported by the user collaboration grant program (UCGP) of the NHMFL which is supported by NSF through NSF-DMR-1157490 and 1644779, and the State of Florida.
J.W. Levitan, J. Lu, B Jarvis, and H. Bai are with National High Magnetic Laboratory, Tallahassee, FL 32310, USA (corresponding author: J. Lu, junlu@ magnet.fsu.edu).

Color versions of one or more of the figures in this paper are available online at http://ieeexplore.ieee.org.
Digital Object Identifier will be inserted here upon acceptance.

total of 40 μm copper stabilizer. The copper tape is 125 μm thick made of oxygen-free copper. The contact area was 25 x 4 mm$^2$.

The vacuum impregnation with paraffin was at about 100 °C. The process took about 1 hour. Before and after wax impregnation, the contact resistivity was obtained by four-leads measurement in liquid nitrogen. Current of ±1 A was applied by a Keithley 2400 source-meter to the contact, whose voltage was measured by a Keithley 2010 digital multimeter.

*B. Coil Test*

It is intuitive that the impact of impregnated wax on a NI-coil is influenced by contact pressure. To calculate the contact pressure distribution in an as-wound coil, a computer code was used. This allowed a scale-down test coil to be wound with a winding tension that results in a distributed contact pressure comparable to that of a practical coil. The radial stress of an as-wound 32 T coil with 20 mm inner radius, 70 mm outer radius, and winding tension of 34.5 MPa was calculated and plotted as the dashed line in Fig. 2. The compressive radial stress (contact pressure) increases from zero at the outer radius and reaches a maximum of 22 MPa at about 30 mm radius. A scale-down coil was designed with winding tension of 50 MPa to have a similar peak radial stress. The parameters of this coil are listed in Table I. The stress distribution of this coil was calculated and plotted as the solid line in Fig. 2.

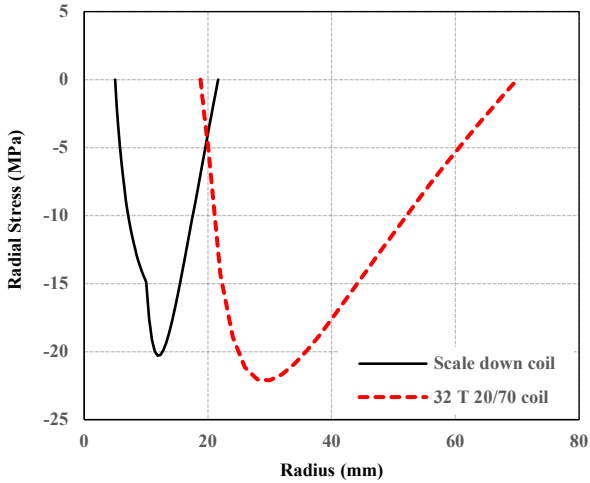

Fig. 2: Comparison of the calculated radial stress on the 32 T Coil A with winding tension of 34.5 MPa and the scale-down test coil with winding tension of 50 MPa.

A G-10 coil mandrel was machined and a single-pancake test coil was wound using SuperPower SCS4050-AP REBCO tape as shown in Figure 3. The wax impregnation process is the same as previously described. For the contact resistivity measurement before and after impregnation, the test coil was immersed in either liquid nitrogen or liquid helium. The coil was charged by a TDK Lambda GEN 8-180 DC power supply to 20 A. The field decay time constant was measured by suddenly discharging the test coil. The field decay was measured by a Hall sensor (HZ-312C by Asahi Kasei Corporation) placed at the bore of the pancake coil. Both current shunt voltage and Hall sensor voltage were measured by a National Instrument SCXI-1000 and recorded by a LabVIEW fast data acquisition software with a time resolution of one data point every 0.1 milliseconds.

TABLE I
COIL PARAMETERS

| Parameters | Values |
|---|---|
| G-10 mandrel ID (mm) | 10 |
| Coil ID (mm) | 20 |
| Coil OD (mm) | 43.3 |
| Number of turns | 128 |
| Winding force (kg) | 2 |

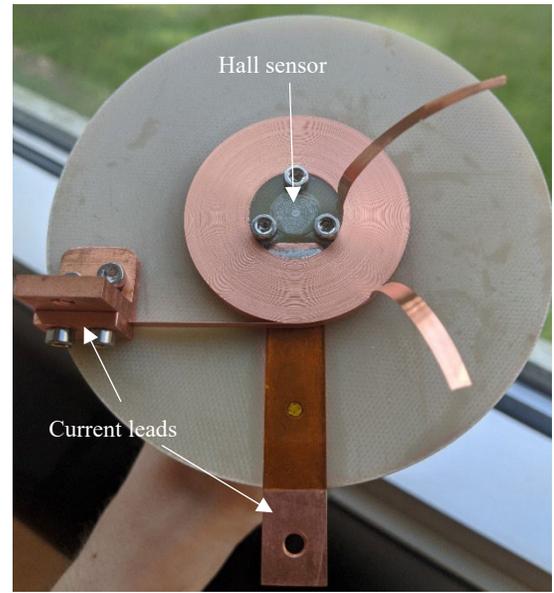

Fig. 3: The single pancake test coil on a G-10 mandrel.

### III. RESULTS AND DISCUSSIONS

*A. Short Sample Test*

Two samples were tested at two different pressures. $R_{ct}$ results are shown in Table II. Evidently, the $R_{ct}$ increases with decreasing contact pressure for both pre- and post-wax cases. $R_{ct}$ seems to increase by wax impregnation. But the increase is moderate, especially for the sample with higher contact pressure.

*B. Coil Test*

The field decay after sudden coil discharge of the wax impregnated coil at 4.2 K are shown together with the exponential decay simulation in Figure 4. The current shunt voltage trace indicates a sufficiently fast switching off, ensuring the validity of this test.

TABLE II
SHORT SAMPLE TEST: $R_{ct}$ AT 77 K, WITH COPPER CO-WIND LAYER

| Sample ID | Pressure, MPa (Torque, in-lb) | Pre-wax $R_{ct}$, μΩ-cm² | Post-wax $R_{ct}$, μΩ-cm² |
|---|---|---|---|
| Rc-248 | ~20 (10) | 95 | 116 |
| Rc-256 | ~10 (5) | 296 | 400 |

TABLE III
COIL TEST RESULTS

| Coil Condition | T (K) | τ (s) | Contact resistivity (μΩ-cm²) |
|---|---|---|---|
| Prewax | 77 | 0.42 | 42 |
| Postwax | 77 | 0.42 | 42 |
| Prewax | 4.2 | 0.68 | 26 |
| Postwax | 4.2 | 0.71 | 25 |

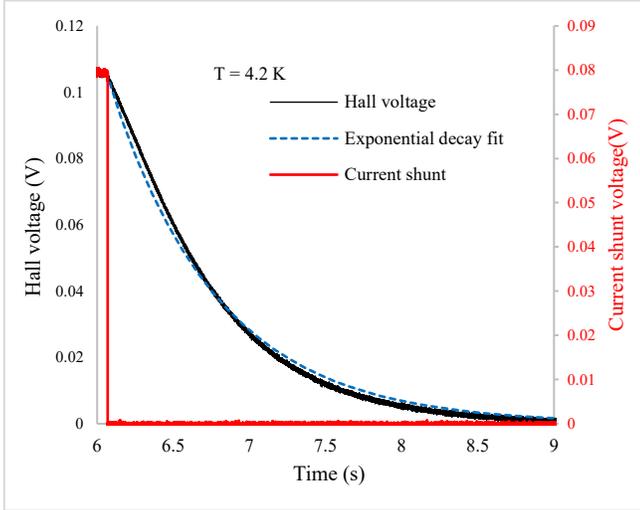

Fig. 4. Data from the sudden coil discharge of the wax impregnated coil at 4.2 K. The black line is the simulation by an exponential decay. The current shunt voltage trace indicates a sufficiently fast switching off.

The decay time constant τ can be readily obtained from the simulated exponential decay function. Subsequently, the contact resistivity $R_{ct}$ can be calculated by

$$R_c = L/\tau \quad (1)$$
$$R_{ct} = R_c \times A/N \quad (2)$$

Where $L$ = 532 μH is the calculated inductance of this coil. τ is the field decay time constant. $R_c$ is the total effective contact resistance of the coil. $A$ is the average turn to turn contact area. $N$ is the total number of turns.

The same coil was first used for the 77 K test, then the 4.2 K test without warming up to room temperature. After the 4.2 K tests, the coil is unwound and checked by scraping the conductor surface with a razor blade. White color wax is observed on conductor surface indicating that the coil is properly impregnated and wax is indeed filled between turns. The decay time constant and calculated contact resistivity are listed in Table III.

## IV. DISCUSSIONS

In the short sample test, there is a copper co-wind tape between two REBCO tapes. In other words, the result is a sum of resistances from two REBCO/Cu interfaces. Therefore, to compare the short sample test results with that of the test coil, its $R_{ct}$ value should be divided by two. This gives us about 48 μΩ-cm² per interface at 20 MPa pressure, and 148 μΩ-cm² at 10 MPa which compares reasonably well with the coil test results of 42 μΩ-cm² at 77 K, although the contact pressure in the coil is distributed from 1 – 20 MPa. Given the significant uncertainty in short sample $R_{ct}$ from sample to sample, the agreement between the short sample and the coil is satisfactory. These values are also in reasonable agreement with what has been reported in the literature [3], [10].

The apparent increase in $R_{ct}$ after wax impregnation in Table II can be partly attributed to the thermal cycling effect of the test. This effect was also observed in two consecutive tests of the same sample before its wax impregnation. In addition, there is always possibility of sample movement during testing and wax impregnation process, which could increase contact resistance, even though Belleville washers were used to keep the contact pressure constant during sample handling.

In the test coil, the contact pressure is a function of radial position of the coil as predicted in Fig. 2. Since $R_{ct}$ is a strong function of contact pressure, it is expected that the $R_{ct}$ also varies with radial position. Obviously, equations (1) and (2) which uses an average contact resistance to describe the field decay process is overly simplified. This can explain the fact that the single exponential fit of experimental data (Fig. 4) is not perfect. For engineering purposes, however, this simple model seems to be adequate in most cases.

Table III shows that the change in $R_{ct}$ of the test coil after wax impregnation is very small at both 77 K and 4.2 K. Since the contact resistivity is determined by the number density and size of the asperity spots [19], the experimental results suggest that the filling of liquid wax into the voids at the interface does not alter the size and density of the asperity spots. The reduction of $R_{ct}$ from 77 K to 4.2 K is likely due to the change in copper resistivity. Similar changes have been reported in [3].

Although our experiments did not indicate any significant effect of wax on contact resistivity, it is still important to discuss the general effect of a liquid in an electrical contact. Because the understanding of the effect of a non-conductive liquid to an electrical contact in general can help to avoid potential pitfalls in an impregnation or a wet winding process. It would be especially important if epoxy impregnation is considered for a NI coil, which, once developed, would significantly improve coil stability against electromagnetic forces [11] and mitigate the issue of $R_{ct}$ drop with pressure cycling [4].

Existence of a non-conductive liquid at the contacting interface does not necessarily influence the contact resistance. For instance, non-conductive liquid lubricants are commonly used in electrical pressure contacts such as switches [12]. In a NI coil, the turn-to-turn contacts are initially made by contact pressure due to coil winding tension. Subsequently in the impregnation process, the liquid wax fills all the voids between turns by capillary effect. If the capillary pressure was comparable with the contact pressure, the liquid would get in between the contacting asperity spots, and significantly increase the contact resistivity. The capillary pressure $P_c$ can be written as

$$P_c = 2\gamma cos\theta/r_c \qquad (3)$$

Where $\gamma$ is the surface tension of the liquid, $\theta$ the wetting angle, and $r_c$ the radius of the capillary. For the paraffin wax, the surface tension at the impregnation temperature of 100 °C is about 28 mN/m [13]. We assume very good wetting with $\theta$ of near 0°, and the capillary $r_c$ of 0.5 μm (the measured roughness of SuperPower REBCO tape surface). Under these assumptions, $P_c$ is in the order of 0.1 MPa, much smaller than the contact pressure in both the short samples and the coil which has distributed contact pressure of 0.5 – 20 MPa. It means that the capillary pressure is not sufficient to drive wax in between contacting spots. Therefore, wax impregnation will not change the contact resistivity significantly. During the wax solidification and eventual cooling down to cryogenic temperatures, the thermal contraction of the wax is considerably more than that of REBCO. So the contact resistivity is unlikely to change due to differential thermal contractions.

Since wax impregnation does not change $R_{ct}$ significantly, it seems to be a safe choice for improving the mechanical stability for an NI coil. The same is true for other pressure contact situations in insulated coils, such as coils wound with multiple REBCO tapes or with a cowind copper tape for stabilization. However, wax impregnation does not provide robust mechanical support to mitigate the issue of remarkable $R_{ct}$ reduction under cyclic pressure due to wearing [5]. A contact of REBCO/stainless-steel which was vacuum impregnated by paraffin experienced remarkable $R_{ct}$ reduction after 30,000 pressure cycles at 4.2 K. It is similar to the case without wax impregnation. Similar to wax impregnation, epoxy impregnation of a NI coil is also unlikely to change the contact resistivity, giving the fact that surface tension and thermal contraction of most epoxy resins are comparable to that of paraffin [14]. In fact, our preliminary experiment in one sample showed no significant increase in the contact resistivity by applying wet Stycast L28 epoxy between REBCO and stainless-steel tape and cured under contact pressure. In addition, this sample was tested under cyclic pressure of 2.5 – 25 MPa up to 30,000 cycles at 4.2 K. The reduction in $R_{ct}$ was only about a factor of 2 compared with a factor of about 1000 in a sample without epoxy [5]. In the case of epoxy impregnated REBCO NI coil, however, a method mitigating the issue of REBCO degradation by epoxy [15] must be developed. Recently, appreciable progress has been made in this area [16]-[18].

## V. Conclusion

In order to determine the effects of wax impregnation on $R_{ct}$ of REBCO tape, a short sample test and a coil test were performed. For the short sample test, an apparatus was designed to compress two pieces of REBCO tape with a copper interlayer at different pressures. Belleville washers were used to hold the pressure during the wax impregnation process. This test showed minimal $R_{ct}$ increase, most likely caused by thermal cycling. A contact resistivity test coil was designed by mimicking the contact pressure of a real coil and a scaled coil was wound with no interlayer. The field decay time constant of the coil was tested at 77 K and 4.2 K before and after wax impregnation. The results from the short sample tests and the coil tests are consistent, which indicate that wax impregnation does not cause significant changes in $R_{ct}$. Wax impregnation can be used to improve mechanical stability of NI coils or coils wound by multiple-tape conductor where low contact resistivity between adjacent tapes is desirable.

## Acknowledgment

This work was performed at the National High Magnetic Field Laboratory, which is supported by National Science Foundation Cooperative Agreement No. DMR-1644779 and DMR-1938789 and the State of Florida. We would like to thank the Applied Superconductivity Center for use of a coil winding machine.